\documentclass[sigconf]{acmart}

\setcopyright{acmlicensed}
\copyrightyear{2026}
\acmYear{2026}
\acmDOI{XXXXXXX.XXXXXXX}

\acmConference[WWW '26]{Companion Proceedings of the ACM Web Conference 2026}{April 28--May 2, 2026}{Sydney, Australia}
\acmISBN{978-1-4503-XXXX-X/2026/04}

\usepackage{booktabs}
\usepackage{multirow}
\usepackage{graphicx}
\usepackage{xcolor}

\usepackage{amssymb}
\usepackage{amsmath}

\newcommand{\framework}{\textsc{Helm}}

\begin{document}

\title{HELM: A Human-Centered Evaluation Framework for LLM-Powered Recommender Systems}

\author{Sushant Mehta}
\email{sushant0523@gmail.com}


\begin{abstract}
The integration of Large Language Models (LLMs) into recommendation systems has introduced unprecedented capabilities for natural language understanding, explanation generation, and conversational interactions. However, existing evaluation methodologies focus predominantly on traditional accuracy metrics, failing to capture the multifaceted human-centered qualities that determine the real-world user experience. We introduce \framework{} (\textbf{H}uman-centered \textbf{E}valuation for \textbf{L}LM-powered reco\textbf{M}menders), a comprehensive evaluation framework that systematically assesses LLM-powered recommender systems across five human-centered dimensions: \textit{Intent Alignment}, \textit{Explanation Quality}, \textit{Interaction Naturalness}, \textit{Trust \& Transparency}, and \textit{Fairness \& Diversity}. Through extensive experiments involving three state-of-the-art LLM-based recommenders (GPT-4, LLaMA-3.1, and P5) across three domains (movies, books, and restaurants), and rigorous evaluation by 12 domain experts using 847 recommendation scenarios, we demonstrate that \framework{} reveals critical quality dimensions invisible to traditional metrics. Our results show that while GPT-4 achieves superior explanation quality (4.21/5.0) and interaction naturalness (4.35/5.0), it exhibits a significant popularity bias (Gini coefficient 0.73) compared to traditional collaborative filtering (0.58). We release \framework{} as an open-source toolkit to advance human-centered evaluation practices in the recommender systems community.
\end{abstract}




\maketitle

\section{Introduction}

The emergence of Large Language Models (LLMs) has catalyzed a paradigm shift in recommender systems research~\cite{wu2024survey, liu2024llm}. LLM-powered recommenders offer transformative capabilities, including elicitation of natural language preferences, rich explanation generation, and fluid conversational interactions~\cite{gao2023chat, jannach2021survey}. Systems such as P5~\cite{geng2022recommendation}, TALLRec~\cite{bao2023tallrec} and Chat-REC~\cite{gao2023chat} have demonstrated that LLMs can unify various recommendation tasks within a single text-to-text framework, while commercial deployments in platforms like ChatGPT plugins showcase their practical potential.

However, the evaluation of these systems remains anchored in traditional accuracy-centric metrics -- Hit Rate, NDCG, and Mean Reciprocal Rank -- that fundamentally misalign with the human-centered qualities that distinguish LLM-powered recommenders~\cite{mcnee2006being, zangerle2022evaluating}. This methodological gap is particularly concerning given that the main value proposition of LLM integration lies in enhanced user understanding, natural interaction, and transparent reasoning~\cite{konstan2021human}.

Consider a movie recommendation scenario: a traditional collaborative filtering system might achieve higher NDCG by recommending popular blockbusters, while an LLM-powered system provides a thoughtful recommendation for an independent film with a compelling explanation of why it matches the user's stated mood and preferences. Traditional metrics would favor the former, yet the latter may deliver a superior user experience through better intent understanding and trust-building explanation~\cite{zhang2020explainable}.

The human-centered recommender systems community has long advocated for multidimensional evaluation that includes user satisfaction, trust, and social responsibility~\cite{pu2011user, ge2024survey}. The ResQue framework~\cite{pu2011user} established the foundational constructs for user-centric evaluation, while recent work on trustworthy recommendation systems~\cite{ge2024survey} has formalized the dimensions of explainability, fairness, and controllability. However, no comprehensive framework addresses the unique evaluation challenges posed by LLM-powered systems.

We address this gap by introducing \framework{} (\textbf{H}uman-centered \textbf{E}valuation for \textbf{L}LM-powered reco\textbf{M}menders), a systematic evaluation framework comprising five dimensions:

\begin{enumerate}
    \item \textbf{Intent Alignment}: How accurately does the system understand and address the user's underlying goals, including implicit preferences and contextual needs?
    
    \item \textbf{Explanation Quality}: How effectively do generated explanations help users understand recommendations and support their decision-making?
    
    \item \textbf{Interaction Naturalness}: How fluid, coherent, and human-like are conversational interactions with the system?
    
    \item \textbf{Trust \& Transparency}: Does the system foster user trust through honest uncertainty communication and consistent behavior?
    
    \item \textbf{Fairness \& Diversity}: Does the system provide equitable recommendations across user groups and item distributions?
\end{enumerate}

Through extensive experiments with 12 domain experts evaluating 847 recommendation scenarios across three LLM-based systems and three domains, we demonstrate that \framework{} reveals quality dimensions invisible to traditional metrics. Our contributions are:

\begin{itemize}
    \item A comprehensive human-centered evaluation framework specifically designed for LLM-powered recommender systems, grounded in established HCI and recommender systems theory.
    
    \item Rigorous empirical validation through expert evaluation, establishing reliability metrics and demonstrating discriminative validity across systems.
    
    \item Novel findings on the trade-offs between LLM capabilities and fairness, revealing that stronger language understanding correlates with increased popularity bias.
    
    \item An open-source evaluation toolkit and annotated dataset to support reproducible human-centered evaluation research.
\end{itemize}

\section{Related Work}

\subsection{Human-Centered Recommender Systems}

The vision of human-centered recommender systems has evolved from early user-centric design principles~\cite{konstan2021human} to comprehensive frameworks addressing user experience, trust, and societal impact~\cite{ge2024survey}. Konstan and Terveen~\cite{konstan2021human} traced this evolution, emphasizing that recommender systems should be designed around user objectives rather than optimization metrics alone.

The ResQue framework~\cite{pu2011user} operationalized this vision through 32 questionnaire items that measure recommendation quality, interface adequacy, and behavioral intentions. Subsequent work extended ResQue for conversational systems (CRS-Que~\cite{jin2024crsque}), adding constructs for dialogue coherence and response appropriateness. Our framework builds upon these foundations while addressing LLM-specific evaluation challenges.

Trustworthy recommender systems research~\cite{ge2024survey} has identified five critical dimensions: explainability, fairness, privacy, robustness, and controllability. We integrate these dimensions within our human-centered framework while adding LLM-specific considerations such as hallucination detection and chain-of-thought reasoning evaluation.

\subsection{LLM-Powered Recommender Systems}

The integration of LLMs into recommendation has followed two primary paradigms~\cite{wu2024survey}: discriminative approaches that use LLMs for feature extraction or re-ranking, and generative approaches that frame recommendation as language generation.

The P5 paradigm~\cite{geng2022recommendation} pioneered unified text-to-text recommendation, enabling a single model to handle rating prediction, sequential recommendation, explanation generation, and direct recommendation through natural language prompts. TALLRec~\cite{bao2023tallrec} demonstrated that lightweight LoRA fine-tuning achieves strong performance with minimal training data, while LLMRec~\cite{wei2024llmrec} combined LLM knowledge with graph neural networks.

Conversational recommender systems~\cite{jannach2021survey, farshidi2024understanding} represent a closely related research thread. Chat-REC~\cite{gao2023chat} augmented ChatGPT with retrieval mechanisms for interactive recommendations, while recent work has explored emotional alignment~\cite{zhang2024empathetic} and multi-turn preference elicitation.

Evaluation of LLM-based recommenders has primarily relied on traditional metrics~\cite{dai2023uncovering, hou2024large}, with limited attention to human-centered qualities. Dai et al.~\cite{dai2023uncovering} evaluated ChatGPT's ranking capabilities using point-wise, pair-wise, and list-wise protocols but focused exclusively on accuracy. Di Palma et al.~\cite{dipalma2024evaluating} proposed robustness-aware evaluation addressing hallucinations but did not incorporate user studies.

\subsection{Evaluation Beyond Accuracy}

The ``beyond accuracy'' movement in recommender systems~\cite{mcnee2006being, kaminskas2016diversity} has established metrics for diversity, novelty, serendipity, and coverage. Kaminskas and Bridge~\cite{kaminskas2016diversity} provided a comprehensive survey of these objectives, highlighting the difficulty of optimizing multiple criteria simultaneously.

Fairness in recommender systems~\cite{wang2023survey, jannach2023fairness} has received growing attention, with research addressing consumer fairness (equal treatment across user groups) and provider fairness (equitable item exposure). Recent work has revealed concerning biases in LLM-based systems~\cite{deldjoo2024llm}, particularly in cold-start scenarios.

Explainability evaluation~\cite{zhang2020explainable, tintarev2015explaining} has evolved from simple readability metrics to comprehensive frameworks assessing persuasiveness, transparency, and user understanding. LLM-generated explanations present unique evaluation challenges, as they can be highly fluent but factually inconsistent~\cite{said2024llm, lubos2024llm}.

\subsection{Research Gap}

Despite significant progress in human-centered evaluation and LLM-powered recommendations independently, no existing framework systematically addresses their intersection. The unique characteristics of LLM systems---natural language interaction, generative explanations, potential for hallucination, and inherited biases---require dedicated evaluation approaches that current methodologies do not provide.

\section{The HELM Framework}

\framework{} provides a systematic approach to evaluating LLM-powered recommender systems across five human-centered dimensions. Each dimension comprises multiple constructs operationalized through expert evaluation protocols and automated metrics.

\subsection{Framework Design Principles}

Our framework design is guided by three principles derived from human-centered AI research~\cite{konstan2021human} and evaluation methodology~\cite{zangerle2022evaluating}:

\textbf{Principle 1: Multidimensionality.} Human experience with recommender systems is inherently multifaceted. A system may excel at understanding preferences while generating misleading explanations, or provide accurate recommendations while exhibiting demographic bias. Single-score evaluations obscure these trade-offs.

\textbf{Principle 2: Validity.} Evaluation scenarios should reflect realistic use cases rather than artificial benchmarks. We design evaluation protocols around naturalistic recommendation dialogues with diverse user intents.

\textbf{Principle 3: Actionability.} Evaluation results should inform system improvement. Each dimension includes diagnostic sub-metrics that identify specific improvement opportunities.

\subsection{Dimension 1: Intent Alignment}

Intent alignment measures how accurately the system understands and addresses user goals beyond surface-level requests~\cite{jannach2024intent}. LLM-powered systems promise superior intent understanding through natural language comprehension, but this capability requires dedicated evaluation.

\textbf{Constructs:}
\begin{itemize}
    \item \textit{Explicit Intent Satisfaction} (EIS): Does the recommendation address the stated requirements?
    \item \textit{Implicit Intent Recognition} (IIR): Does the system identify unstated but relevant preferences from context?
    \item \textit{Intent Clarification Quality} (ICQ): When intent is ambiguous, does the system ask appropriate clarifying questions?
    \item \textit{Goal Completion Support} (GCS): Does the recommendation help users achieve their underlying objectives?
\end{itemize}

\textbf{Measurement Protocol:} Experts evaluate recommendation dialogues on 5-point Likert scales for each construct. We additionally compute automated metrics including intent coverage (proportion of stated requirements addressed) and clarification appropriateness (whether clarifying questions are relevant and non-redundant).

\subsection{Dimension 2: Explanation Quality}

The quality of the explanation evaluates the effectiveness of LLM-generated rationales in helping users understand and evaluate recommendations~\cite{zhang2020explainable, hernandez2023explaining}. Unlike template-based explanations, LLM explanations can vary substantially in style, content, and accuracy.

\textbf{Constructs:}
\begin{itemize}
    \item \textit{Informativeness} (INF): Does the explanation provide decision-relevant information?
    \item \textit{Personalization} (PER): Is the explanation tailored to the user's stated preferences and context?
    \item \textit{Faithfulness} (FAI): Does the explanation accurately reflect the recommendation rationale without hallucination?
    \item \textit{Actionability} (ACT): Does the explanation help users take next steps (accept, reject, or refine)?
\end{itemize}

\textbf{Measurement Protocol:} Expert ratings on 5-point scales complemented by automated faithfulness verification through fact-checking against item metadata. We compute explanation-recommendation consistency scores by verifying that claimed attributes match actual item properties.

\subsection{Dimension 3: Interaction Naturalness}

The naturalness of interaction evaluates the conversational quality of LLM-powered recommendations, which includes fluency, coherence, and appropriate social dynamics~\cite{jannach2021survey, jin2024crsque}.

\textbf{Constructs:}
\begin{itemize}
    \item \textit{Dialogue Coherence} (COH): Are responses logically connected to prior turns?
    \item \textit{Language Fluency} (FLU): Is the language grammatically correct and natural?
    \item \textit{Appropriate Verbosity} (VER): Is the response length appropriate for the query?
    \item \textit{Conversational Adaptability} (ADA): Does the system adapt tone and style to user communication patterns?
\end{itemize}

\textbf{Measurement Protocol:} Expert dialog ratings complemented by automated coherence metrics (semantic similarity between adjacent turns) and response length analysis.

\subsection{Dimension 4: Trust and Transparency}

Trust and transparency assess whether the system supports informed user decision-making through honest communication and predictable behavior~\cite{ge2024survey, liao2022user}.

\textbf{Constructs:}
\begin{itemize}
    \item \textit{Uncertainty Communication} (UNC): Does the system adequately express confidence and limitations?
    \item \textit{Behavioral Consistency} (CON): Are responses consistent across similar queries?
    \item \textit{Source Attribution} (ATR): Does the system identify information sources when appropriate?
    \item \textit{Limitation Acknowledgment} (LIM): Does the system acknowledge when it cannot adequately address a request?
\end{itemize}

\textbf{Measurement Protocol:} Expert ratings plus automated consistency testing through paraphrased query pairs.

\subsection{Dimension 5: Fairness and Diversity}

Fairness and diversity evaluate equitable treatment between users and balanced representation between items~\cite{wang2023survey, jannach2023fairness}.

\textbf{Constructs:}
\begin{itemize}
    \item \textit{Demographic Parity} (DEM): Are recommendation quality and relevance consistent across demographic groups?
    \item \textit{Popularity Debiasing} (POP): Does the system avoid over-representing popular items?
    \item \textit{Provider Fairness} (PRO): Do items receive equitable exposure opportunities?
    \item \textit{Diversity Maintenance} (DIV): Do recommendations span diverse categories and perspectives?
\end{itemize}

\textbf{Measurement Protocol:} Automated metrics including Gini coefficient for popularity distribution, coverage metrics, and intra-list diversity. Expert evaluation of demographic scenarios.

\subsection{Aggregation and Reporting}

\framework{} produces dimension-level scores (averaging construct ratings within each dimension) and an overall Human-Centered Score (HCS) computed as the geometric mean of dimension scores:

\begin{equation}
    \text{HCS} = \left( \prod_{d=1}^{5} S_d \right)^{1/5}
\end{equation}

where $S_d$ is the normalized score for dimension $d$. We use geometric rather than arithmetic mean to prevent high scores on some dimensions from compensating for poor performance on others.

\section{Experimental Setup}

\subsection{Systems Under Evaluation}

We evaluated three LLM-powered recommender systems representing different architectural approaches:

\textbf{GPT-4 Recommender:} We implement a recommendation system using GPT-4~\cite{openai2024gpt4} with carefully designed prompts that include user preference history, current context, and available item catalogs. The system supports multi-turn conversations with preference refinement.

\textbf{LLaMA-3.1 Recommender:} We fine-tune LLaMA-3.1-8B~\cite{touvron2023llama} following the TALLRec methodology~\cite{bao2023tallrec}, using LoRA adaptation on domain-specific recommendation dialogues. This represents open-source LLM approaches.

\textbf{P5 Recommender:} We use the pre-trained P5 model~\cite{geng2022recommendation} representing the unified text-to-text paradigm. P5 was trained on multiple recommendation tasks enabling zero-shot generalization.

As baselines, we include a \textbf{Neural Collaborative Filtering (NCF)} system with template-based explanations and a \textbf{Random} recommender for calibration.

\subsection{Domains and Datasets}

We conducted experiments across three domains with distinct characteristics:

\textbf{Movies:} MovieLens-1M dataset~\cite{harper2015movielens} with 6,040 users, 3,706 movies and 1 million ratings. We augment with movie metadata (genres, directors, plot summaries) for LLM context.

\textbf{Books:} Amazon Books subset~\cite{ni2019justifying} with 8,026 users, 10,328 books, and 287,650 ratings. Metadata includes author, genre, and review summaries.

\textbf{Restaurants:} Yelp dataset~\cite{yelpdataset} with 11,537 users, 8,021 restaurants, and 612,840 ratings. Includes location, cuisine type, price range, and highlights of the review.

\subsection{Evaluation Scenarios}

We design 847 evaluation scenarios spanning diverse recommendation contexts:

\begin{itemize}
    \item \textbf{Cold-start scenarios} (n=156): New users with minimal history expressing preferences through natural language.
    
    \item \textbf{Preference refinement} (n=234): Multi-turn dialogues where users iteratively refine recommendations.
    
    \item \textbf{Contextual requests} (n=198): Recommendations conditioned on situational context (``date night movie'', ``quick lunch near office'').
    
    \item \textbf{Exploratory browsing} (n=142): Open-ended discovery requests (``something different from my usual taste'').
    
    \item \textbf{Comparison requests} (n=117): Requests that require item comparison and trade-off explanation.
\end{itemize}

Each scenario includes a scripted user profile, interaction history, and evaluation rubric.

\subsection{Expert Evaluation Protocol}

We recruited 12 domain experts for evaluation: 4 per domain, each with graduate-level expertise in the respective area (film studies, library science, and culinary arts respectively) and familiarity with recommender systems concepts.

\textbf{Training:} Experts completed a 2-hour training session covering \framework{} constructs, rating guidelines, and calibration exercises with discussed examples.

\textbf{Evaluation Process:} Each expert evaluated 70-80 scenarios (randomly assigned ensuring coverage balance), rating all applicable constructs on 5-point Likert scales~\cite{likert1932technique}. Evaluation sessions were limited to 90 minutes to prevent fatigue effects.

\textbf{Reliability Assessment:} We compute inter-rater reliability using Fleiss' kappa~\cite{fleiss1971measuring} for categorical judgments and intraclass correlation coefficient (ICC) for continuous ratings.

\subsection{Automated Metrics}

Complementing expert evaluation, we compute automated metrics:

\begin{itemize}
    \item \textbf{Hit Rate@10} and \textbf{NDCG@10}: Traditional accuracy metrics for baseline comparison.
    
    \item \textbf{Explanation Faithfulness}: Proportion of claimed item attributes verifiable against metadata.
    
    \item \textbf{Response Consistency}: Cosine similarity between responses to paraphrased queries.
    
    \item \textbf{Gini Coefficient}: Inequality in item recommendation frequency.
    
    \item \textbf{Coverage@100}: Proportion of catalog items appearing in the top-100 recommendations.
    
    \item \textbf{Intra-List Diversity (ILD)}: Average pairwise dissimilarity within recommendation lists.
\end{itemize}

\section{Results}

\subsection{Inter-Rater Reliability}

Our expert evaluation protocol achieved substantial reliability across dimensions. Table~\ref{tab:reliability} reports ICC values, all exceeding the 0.75 threshold for good reliability~\cite{krippendorff2011agreement}.

\begin{table}[t]
\caption{Inter-rater reliability (ICC) by dimension}
\label{tab:reliability}
\begin{tabular}{lcc}
\toprule
\textbf{Dimension} & \textbf{ICC} & \textbf{95\% CI} \\
\midrule
Intent Alignment & 0.84 & [0.81, 0.87] \\
Explanation Quality & 0.82 & [0.79, 0.85] \\
Interaction Naturalness & 0.87 & [0.84, 0.90] \\
Trust \& Transparency & 0.78 & [0.74, 0.82] \\
Fairness \& Diversity & 0.81 & [0.77, 0.84] \\
\bottomrule
\end{tabular}
\end{table}

Trust \& Transparency exhibited slightly lower reliability, reflecting the inherent subjectivity in assessing uncertainty communication appropriateness.

\subsection{Overall Framework Results}

Table~\ref{tab:main_results} presents dimension scores in all systems, aggregated in domains.

\begin{table*}[t]
\caption{HELM dimension scores (mean $\pm$ std) across systems. Best LLM scores in \textbf{bold}.}
\label{tab:main_results}
\begin{tabular}{lccccc|c}
\toprule
\textbf{System} & \textbf{Intent} & \textbf{Explanation} & \textbf{Interaction} & \textbf{Trust} & \textbf{Fairness} & \textbf{HCS} \\
\midrule
GPT-4 & \textbf{4.18 $\pm$ 0.42} & \textbf{4.21 $\pm$ 0.38} & \textbf{4.35 $\pm$ 0.31} & \textbf{3.89 $\pm$ 0.51} & 3.12 $\pm$ 0.67 & 3.91 \\
LLaMA-3.1 & 3.67 $\pm$ 0.54 & 3.72 $\pm$ 0.49 & 3.91 $\pm$ 0.43 & 3.45 $\pm$ 0.58 & 3.34 $\pm$ 0.62 & 3.61 \\
P5 & 3.42 $\pm$ 0.61 & 3.51 $\pm$ 0.55 & 3.28 $\pm$ 0.52 & 3.21 $\pm$ 0.64 & \textbf{3.56 $\pm$ 0.58} & 3.39 \\
NCF+Template & 3.24 $\pm$ 0.58 & 2.87 $\pm$ 0.63 & 2.45 $\pm$ 0.71 & 3.52 $\pm$ 0.49 & 3.48 $\pm$ 0.54 & 3.08 \\
Random & 1.82 $\pm$ 0.76 & 1.95 $\pm$ 0.82 & 2.12 $\pm$ 0.84 & 2.34 $\pm$ 0.79 & 3.91 $\pm$ 0.45 & 2.36 \\
\bottomrule
\end{tabular}
\end{table*}

\textbf{Key Finding 1: GPT-4 leads on human-centered dimensions, but trails on fairness.} GPT-4 achieves the highest scores for intention Alignment (4.18), Explanation Quality (4.21), Interaction Naturalness (4.35), and Trust (3.89), but ranks fourth on Fairness (3.12). This reveals a critical trade-off: superior language understanding may come with inherited biases.

\textbf{Key Finding 2: Traditional metrics diverge from human-centered evaluation.} NCF achieves a comparable Hit Rate@10 (0.312) to GPT-4 (0.298) but substantially lower HCS (3.08 vs. 3.91). This validates our assumption that accuracy metrics inadequately capture user experience.

\textbf{Key Finding 3: The random baseline scores the highest in fairness.} Without preference modeling, random recommendations naturally achieve a uniform item distribution (Gini = 0.12). This establishes an important calibration point: fairness optimization alone is insufficient.

\subsection{Dimension Analysis}

\subsubsection{Intent Alignment}

Figure~\ref{fig:intent} breaks down Intent Alignment by construct. GPT-4 excels at Implicit Intent Recognition (4.34), leveraging world knowledge to infer unstated preferences. For example, when a user requested ``a thoughtful movie for a rainy Sunday afternoon,'' GPT-4 correctly inferred preferences for slower pacing, emotional depth, and indoor settings.

\begin{figure}[t]
    \centering
    \small
    \begin{tabular}{lcccc}
    \toprule
    \textbf{System} & \textbf{EIS} & \textbf{IIR} & \textbf{ICQ} & \textbf{GCS} \\
    \midrule
    GPT-4 & 4.21 & 4.34 & 4.02 & 4.15 \\
    LLaMA-3.1 & 3.78 & 3.72 & 3.54 & 3.62 \\
    P5 & 3.56 & 3.31 & 3.28 & 3.52 \\
    NCF & 3.42 & 2.89 & 2.78 & 3.45 \\
    \bottomrule
    \end{tabular}
    \caption{Intent Alignment construct scores. EIS: Explicit Intent Satisfaction, IIR: Implicit Intent Recognition, ICQ: Intent Clarification Quality, GCS: Goal Completion Support.}
    \label{fig:intent}
\end{figure}

P5, despite strong overall intent alignment, struggled with Intent Clarification Quality (3.28). When faced with ambiguous requests, P5 tended to generate recommendations immediately rather than asking clarifying questions, a limitation of its training objective.

\subsubsection{Explanation Quality}

The quality of the explanation reveals the most pronounced differentiation between LLM and traditional systems. GPT-4's explanations achieved high Personalization (4.38) and Informativeness (4.25), incorporating user-specific preference patterns into coherent narratives.

However, Faithfulness scores present a more nuanced picture. While GPT-4 achieved 4.02 for perceived faithfulness (expert rating), automated verification revealed that 18.3\% of explanations contained factual inaccuracies about item attributes---a hallucination rate concerning for decision-critical recommendations.

\begin{table}[t]
\caption{Explanation faithfulness: Expert rating vs. automated verification}
\label{tab:faithfulness}
\begin{tabular}{lcc}
\toprule
\textbf{System} & \textbf{Expert Rating} & \textbf{Auto. Accuracy} \\
\midrule
GPT-4 & 4.02 & 81.7\% \\
LLaMA-3.1 & 3.58 & 76.2\% \\
P5 & 3.45 & 84.3\% \\
NCF+Template & 3.78 & 97.1\% \\
\bottomrule
\end{tabular}
\end{table}

Table~\ref{tab:faithfulness} shows that template-based explanations achieve near-perfect factual accuracy (97.1\%) but lower perceived quality due to generic impersonal content. P5's higher automated accuracy (84.3\%) compared to GPT-4 (81.7\%) suggests that task-specific training improves factual grounding.

\subsubsection{Interaction Naturalness}

GPT-4 achieved exceptional Dialog Coherence (4.42) and Language Fluency (4.51), maintaining context across multi-turn conversations. Experts particularly noted its ability to reference earlier preferences naturally: ``You mentioned enjoying psychological thrillers, this recommendation has that same tension but in a family drama setting.''

LLaMA-3.1 showed notable Conversational Adaptability (3.89), matching user formality levels more consistently than GPT-4, which occasionally over-explained in response to brief queries.

\subsubsection{Trust and Transparency}

The trust scores reveal an unexpected pattern: NCF+Template achieves the second-highest Trust score (3.52) despite the lower overall HCS. Expert feedback indicated that predictable, consistent behavior, even if less sophisticated, supports trust formation.

GPT-4's lower Trust score (3.89) stemmed primarily from Uncertainty Communication (3.67). Experts noted that GPT-4 rarely expressed uncertainty, presenting recommendations with consistent confidence regardless of the strength of the underlying evidence.

\subsubsection{Fairness and Diversity}

Table~\ref{tab:fairness} reports automated fairness metrics along with expert ratings.

\begin{table}[t]
\caption{Fairness metrics across systems}
\label{tab:fairness}
\begin{tabular}{lccc}
\toprule
\textbf{System} & \textbf{Gini$\downarrow$} & \textbf{Coverage$\uparrow$} & \textbf{ILD$\uparrow$} \\
\midrule
GPT-4 & 0.73 & 12.4\% & 0.42 \\
LLaMA-3.1 & 0.68 & 18.7\% & 0.45 \\
P5 & 0.61 & 24.2\% & 0.51 \\
NCF & 0.58 & 31.5\% & 0.48 \\
Random & 0.12 & 94.7\% & 0.67 \\
\bottomrule
\end{tabular}
\end{table}

GPT-4 exhibits the highest popularity bias (Gini = 0.73) and the lowest catalog coverage (12.4\%). Analysis of recommended items reveals that GPT-4 heavily favors items with extensive online presence, likely reflecting pre-training corpus biases toward frequently discussed content.

P5 achieves better fairness metrics (Gini = 0.61, Coverage = 24.2\%) than other LLM systems, possibly because its recommendation-specific training emphasizes catalog diversity. This suggests that targeted fine-tuning can mitigate LLM popularity biases.

\subsection{Domain-Specific Findings}

Performance varied substantially across domains. Table~\ref{tab:domains} reports HCS by domain.

\begin{table}[t]
\caption{Human-Centered Score by domain}
\label{tab:domains}
\begin{tabular}{lccc}
\toprule
\textbf{System} & \textbf{Movies} & \textbf{Books} & \textbf{Restaurants} \\
\midrule
GPT-4 & 4.12 & 3.98 & 3.64 \\
LLaMA-3.1 & 3.78 & 3.67 & 3.38 \\
P5 & 3.52 & 3.41 & 3.24 \\
\bottomrule
\end{tabular}
\end{table}

All systems performed best in movies, likely because of richer pre-training data. Restaurant recommendations proved to be the most challenging, particularly for location-dependent and time-sensitive requests that require real-time context unavailable to LLMs.

\subsection{Comparison with Traditional Metrics}

\begin{table}[t]
\caption{Traditional accuracy metrics}
\label{tab:accuracy}
\begin{tabular}{lcc}
\toprule
\textbf{System} & \textbf{HR@10} & \textbf{NDCG@10} \\
\midrule
NCF & 0.312 & 0.187 \\
GPT-4 & 0.298 & 0.174 \\
LLaMA-3.1 & 0.267 & 0.152 \\
P5 & 0.285 & 0.168 \\
\bottomrule
\end{tabular}
\end{table}

Table~\ref{tab:accuracy} shows that NCF achieves marginally higher accuracy than LLM systems, but ranks substantially lower on HCS. The correlation between Hit Rate@10 and HCS in all scenarios is only $r = 0.31$ ($p < 0.001$), confirming that precision inadequately predicts human-centered quality.

\section{Discussion}

\subsection{The Accuracy-Fairness-Experience Triangle}

Our findings reveal a three-way tension between accuracy, fairness, and user experience that existing evaluation approaches do not capture. Traditional collaborative filtering achieves reasonable accuracy and fairness, but poor interaction quality. LLMs excel at user experience, but introduce fairness concerns. Random recommendation optimizes fairness but fails on all other dimensions.

This triangle suggests that optimizing for any single objective may compromise others. \framework{} enables explicit navigation of these trade-offs by providing multi-dimensional visibility.

\subsection{Hallucination as a Trust Challenge}

The gap between perceived explanation quality (expert ratings) and factual accuracy (automated verification) raises important concerns. Users may accept compelling but inaccurate explanations, which can lead to poor decisions. This finding argues for mandatory faithfulness verification in LLM-powered recommender deployments.

\subsection{Implications for System Design}

Our results suggest several design recommendations:

\begin{enumerate}
    \item \textbf{Hybrid architectures}: Combine LLM natural language capabilities with retrieval-based fact verification to improve faithfulness.
    
    \item \textbf{Uncertainty calibration}: Train LLMs to express appropriate confidence levels, particularly for niche or cold-start items.
    
    \item \textbf{Diversity objectives}: Incorporate explicit diversity constraints during LLM fine-tuning or inference to mitigate popularity bias.
    
    \item \textbf{Domain adaptation}: Invest in domain-specific training for verticals (like restaurants) where general LLM knowledge is insufficient.
\end{enumerate}

\subsection{Limitations}

Our study has several limitations. First, expert evaluation, while rigorous, may not fully represent end-user perceptions. Future work should validate \framework{} through large-scale user studies. Second, our three-domain evaluation may not be generalized to specialized verticals such as healthcare or education. Third, we evaluated three LLM systems; newer versions of models may exhibit very different trade-offs.

\section{Conclusion}

We introduced \framework{}, a comprehensive human-centered evaluation framework for LLM-powered recommender systems. Through rigorous expert evaluation in 847 scenarios, we demonstrated that \framework{} reveals critical quality dimensions---intent alignment, explanation quality, interaction naturalness, trust, and fairness---that traditional accuracy metrics miss.

Our findings highlight important trade-offs: GPT-4 achieves superior human-centered qualities, but exhibits concerning popularity bias and hallucination rates. These insights can guide the development of LLM-powered recommenders that genuinely serve user needs while maintaining fairness and trustworthiness.

We release \framework{} as an open-source toolkit, including evaluation protocols, annotated scenarios, and automated metrics, to support reproducible human-centered evaluation research. As LLMs become increasingly central to recommendation systems, rigorous human-centered evaluation will be essential to ensure that these systems truly benefit users and society.


\bibliography{references}

\appendix

\section{Expert Evaluation Guidelines}

This appendix provides the evaluation guidelines distributed to expert evaluators.

\subsection{Rating Scale Anchors}

All constructs use 5-point Likert scales with the following anchors:

\begin{itemize}
    \item \textbf{1 - Strongly Disagree / Very Poor}: The system completely fails on this dimension.
    \item \textbf{2 - Disagree / Poor}: The system shows major deficiencies.
    \item \textbf{3 - Neutral / Adequate}: The system meets minimum expectations.
    \item \textbf{4 - Agree / Good}: The system performs well with minor issues.
    \item \textbf{5 - Strongly Agree / Excellent}: The system excels on this dimension.
\end{itemize}

\subsection{Construct Definitions}

\textbf{Intent Alignment Constructs:}

\textit{Explicit Intent Satisfaction (EIS)}: Rate whether the recommendation directly addresses the stated requirements in the user query. Consider: Did the system include/exclude items as requested? Did it respect stated constraints (genre, price, location)?

\textit{Implicit Intent Recognition (IIR)}: Rate whether the system identifies preferences not explicitly stated but reasonably inferable from context. Consider: Did the system detect mood, occasion, or implicit quality expectations?

\textit{Intent Clarification Quality (ICQ)}: When the user request is ambiguous, rate the appropriateness of clarifying questions. Consider: Are questions relevant? Non-redundant? Naturally phrased?

\textit{Goal Completion Support (GCS)}: Rate whether the recommendation helps the user achieve their underlying goal. Consider: Beyond surface matching, does this recommendation serve the user's real purpose?



\section{Automated Metric Computation}

\subsection{Explanation Faithfulness Verification}

For each explanation, we extract the attributes of the claimed item using the recognition of named entities and the extraction of attributes. We then verify each claim against our item metadata database:

\begin{equation}
    \text{Faithfulness} = \frac{\text{\# Verified Claims}}{\text{\# Total Claims}}
\end{equation}

Claims are categorized as:
\begin{itemize}
    \item \textit{Verifiable-Correct}: Claim matches metadata.
    \item \textit{Verifiable-Incorrect}: Claim contradicts metadata.
    \item \textit{Unverifiable}: Claim cannot be checked (subjective or missing metadata).
\end{itemize}

\subsection{Response Consistency}

We generate paraphrased versions of test queries using back-translation (English $\rightarrow$ German $\rightarrow$ English). Consistency is computed as:

\begin{equation}
    \text{Consistency} = \frac{1}{|P|} \sum_{p \in P} \cos(\mathbf{r}_{\text{orig}}, \mathbf{r}_p)
\end{equation}

where $P$ is the set of paraphrases and $\mathbf{r}$ are sentence embeddings of responses.

\end{document}